\documentclass[prd,aps,floats,twocolumn]{revtex4}

\usepackage[dvips]{graphicx}
\usepackage{amssymb}
\usepackage{amsmath}

\usepackage{color}

\begin{document}

\title{Maximum black-hole spin from quasi-circular binary mergers}

\author{Michael Kesden} \email{mhk10@nyu.edu}

\affiliation{Center for Cosmology and Particle Physics, New York University,
Department of Physics, 4 Washington Pl., New York, New York 10003}

\author{Guglielmo Lockhart}

\affiliation{California Institute of Technology, MC 350-17, 1216 E. California
  Blvd., Pasadena, California 91125}

\author{E. Sterl Phinney} \email{esp@tapir.caltech.edu}

\affiliation{California Institute of Technology, MC 350-17, 1216 E. California
  Blvd., Pasadena, California 91125}

\date{October 2010}

\begin{abstract}
Black holes of mass $M$ must have a spin angular momentum $S$ below
the Kerr limit $(\chi \equiv S/M^2 \leq 1)$, but whether astrophysical
black holes can attain this limiting spin depends on their accretion
history.  Gas accretion from a thin disk limits the black-hole spin to
$\chi_{\rm gas} \lesssim 0.9980 \pm 0.0002$, as electromagnetic
radiation from this disk with retrograde angular momentum is
preferentially absorbed by the black hole.  Extrapolation of
numerical-relativity simulations of equal-mass binary black-hole
mergers to maximum initial spins suggests these mergers yield a
maximum spin $\chi_{eq} \lesssim 0.95$.  Here we show that for smaller
mass ratios $q \equiv m/M \ll 1$, the superradiant extraction of
angular momentum from the larger black hole imposes a fundamental
limit $\chi_{\rm lim} \lesssim 0.9979 \pm 0.0001$ on the final
black-hole spin even in the test-particle limit $(q \to 0)$ of binary
black-hole mergers.  The nearly equal values of $\chi_{\rm gas}$ and
$\chi_{\rm lim}$ imply that measurement of supermassive black-hole
spins cannot distinguish a black hole built by gas accretion from one
assembled by the gravitational inspiral of a disk of compact stellar
remnants.  We also show how superradiant scattering alters the mass
and spin predicted by models derived from extrapolating test-particle
mergers to finite mass ratios.
\end{abstract}

\maketitle

\section{Introduction} \label{S:intro}

Supermassive black holes (SBHs) reside in the centers of most large
galaxies.  While a few nearby SBHs can be detected by their
gravitational influence on surrounding stars, the majority of SBHs are
observed electromagnetically as active galactic nuclei (AGN).  The
same accretion flows that release energy to power AGN also supply
energy and angular momentum to the black holes themselves increasing
their mass $M$ and spin $S$ \footnote{Throughout this paper we express
physical quantities in units where Newton's constant $G$ and the speed
of light $c$ are unity, in which case the spin angular momentum $S$
has dimensions of $M^2$.}.  By definition, black holes possess an
event horizon from within which nothing can escape to future null
infinity.  Black holes described by the Kerr metric
\cite{Kerr:1963ud} only possess an event horizon for $\chi \equiv
S/M^2 \leq 1$, setting a fundamental upper limit on a black hole's
possible spin.  Whether or not astrophysical SBHs saturate or even
exceed this Kerr limit is an important test of general relativity.

Black-hole spins also probe their assembly history.  SBHs grow both by
gas accretion and mergers driven by the gravitational inspiral of
binary companions.  These two growth mechanisms may supply mass and
angular momentum to the SBHs in different ratios, allowing
measurements of black-hole spin to distinguish between them.
Reverberation mapping of the iron K$\alpha$ line in AGN x-ray spectra
has been proposed as just such a means of measuring SBH spins
\cite{Reynolds:1998ie}.  XMM-Newton observations analyzed with this
technique have been used to constrain the spin of the SBH hosted by
the Seyfert 1.2 galaxy MCG-06-30-15 to $\chi =
0.989_{-0.002}^{+0.009}$ at 90\% confidence \cite{Brenneman:2006hw}.
A closer examination of how SBHs acquire their spins is thus of both
theoretical and observational importance.

According to Bardeen \cite{Bardeen:1970}, a nonspinning black hole
can attain the maximum Kerr spin $\chi = 1$ after accreting a finite
mass of test particles freely falling from its innermost stable
circular orbit (ISCO).  However, material on circular orbits with
radii greater than that of the ISCO cannot be accreted unless it has
some mechanism to shed its excess angular momentum.  In a standard
geometrically thin, optically thick accretion disk
\cite{Shakura:1972te}, this mechanism is viscous stress within the
disk that also heats it locally and transports energy outwards.  This
heating produces an energy flux at the disk's surface that will be
radiated in all directions.  A small fraction of these radiated
photons do not escape to infinity but are instead captured by the
black hole itself.  Photons with negative angular momentum with
respect to the black hole's spin have a larger capture cross-section
than those with positive angular momentum \cite{Godfrey:1970am},
implying that the accreted radiation will counteract the directly
advected material which always acts to spin the hole up for $\chi <
1$.  These two sources of angular momentum cancel for black-hole spins
$\chi_{\rm gas} \simeq 0.998$, with black holes of greater spins
spinning down to this value after accreting a mass $m \simeq 0.05M$
\cite{Thorne:1974ve}.

Different accretion flows will supply energy and angular momentum to
the black hole in different ratios, altering the value of the limiting
spin $\chi_{\rm gas}$.  Advection-dominated accretion flows (ADAFs) do
not cool efficiently, and therefore a fraction $f > 0$ of the
gravitational energy dissipated prior to the ISCO is advected by the
black hole rather than radiated to infinity.  Viscous stresses in
these flows proportional to the Shakura-Sunyaev parameter $\alpha$
\cite{Shakura:1972te} also reduce the specific angular momentum of the
accreted material below its value in the vacuum at the ISCO.  The
former effect increases $M$ in the denominator in the definition $\chi
\equiv S/M^2$ of the dimensionless spin, while the latter effect
reduces $S$ in the numerator \cite{Popham:1998iw}.  The
magnetorotational instability \cite{Balbus:1991ay} also torques
gas at the ISCO, and in addition can launch jets which further limit
the spin as shown in magnetohydrodynamic simulations
\cite{Gammie:2003qi}.  Analytic fits to the ADAF simulations of Popham
and Gammie \cite{Popham:1998iw} suggest that black holes can spin up
to $\chi_{\rm ADAF} \simeq 0.96$, and that the inclusion of jets
calibrated by the magnetohydrodynamic simulations of Gammie {\it et al.}
\cite{Gammie:2003qi} reduces this limit to $\chi_{\rm jets} \simeq 0.93$
\cite{Benson:2009kx}.  We see that limits on black-hole spin depend
greatly on the nature of the accretion flow, and that no
model-independent constraints exclude the limit $\chi_{\rm gas} \simeq
0.998$ for thin disks first obtained by Thorne \cite{Thorne:1974ve}.

The maximum spin of black holes produced in binary mergers is also
greatly uncertain.  The most accurate numerical-relativity simulations
have been of equal-mass $q \equiv m_2/m_1 = 1$ black holes with spins
$\chi_1 = \chi_2$ aligned or antialigned with the orbital angular
momentum.  These simulations found $\chi = 0.68646 \pm 0.00004$ for
nonspinning binaries \cite{Scheel:2008rj}, and $\chi = 0.547812 \pm
0.000009$ for equal-mass binaries with initial spins $\chi_1 = \chi_2
= -0.43757$ \cite{Chu:2009md}.  The spins of such aligned binaries do
not precess prior to merger, removing one complication in determining
their magnitude and direction.  We will therefore restrict our
analysis to aligned binaries in this paper, though with modest
additional work it could be extended to binaries of arbitrarily
oriented spins.

Although these numerically determined spins are remarkably precise,
two significant obstacles still prevent the determination of final
black-hole spins from more generic mergers.  The first is the
inability of the commonly used conformally flat Bowen-York initial
data to adequately approximate initial binary black holes with spins
greater than $\chi_i \simeq 0.93$ \cite{Lovelace:2008tw}.  An
extrapolation of simulations with $\chi_i \leq 0.9$ suggests that the
merger of equal-mass black holes with maximal initial spins aligned
with their orbital angular momentum yields a final black hole with
spin $\chi = 0.951 \pm 0.004$ \cite{Marronetti:2007wz}.  However, it
remains to be proven that such a smooth extrapolation in initial spin
holds all the way to the Kerr limit.

The second obstacle to determining spins from generic mergers is the
increased computational resources needed to simulate mergers with mass
ratios $q \leq 1$.  This increased computational demand results from
the need for more closely spaced numerical grid points to resolve the
horizon of the smaller black hole, shorter timesteps owing to the
shorter light propagation time between grid points, and longer
simulations to capture the larger number of orbits per unit increase
in orbital frequency prior to merger.  The smallest mass ratio that
has currently been simulated is $q = 0.01$ \cite{Lousto:2010ut}; only
a handful of simulated mergers with $q \leq 0.1$, all of initially
nonspinning black holes, have been published
\cite{Gonzalez:2008bi,Lousto:2010tb,Lousto:2010qx}.  New numerical
techniques will be required to make much progress beyond this point.

Relativists have attempted to surmount these obstacles by inventing
fitting formulas that are functions of $q$, $\chi_1$, and $\chi_2$,
calibrating the coefficients in these formulas with existing
simulations, and then extrapolating them to higher spins and lower
mass ratios than can currently be simulated.  This approach is very
effective in the region of the parameter space $\{q, \chi_1, \chi_2
\}$ near the simulations with which the fitting formulas were
calibrated, but can break down outside this region.  The probable
reason for this is that all three parameters listed above equal unity
for equal-mass, maximally spinning, aligned mergers, and thus
polynomials in these parameters will converge slowly if at all.  Even
the symmetric mass ratio $\eta \equiv m_1m_2/(m_1 + m_2)^2 = q/(1 +
q)^2$ gets as large as 1/4 for equal-mass mergers, suggesting that
polynomials in this parameter will not converge quickly either.  At
the time this paper was written, simulations were restricted to the
region $0.1 \leq q \leq 1$ ($0.0826 \leq \eta \leq 0.25$) which
provided a short lever arm over which to calibrate terms with
different $q$ dependence.  The degeneracy between terms makes
predicting final spins in the test-particle ($q \to 0$) limit very
uncertain.  For example, several fitting formulas predict final spins
above the Kerr limit for maximally spinning aligned mergers with mass
ratios as modest as $q \lesssim 0.25$.  The publication of a
simulation with $q = 0.01$ \cite{Lousto:2010ut} during the preparation
of this paper shows the rapid progress towards numerical simulations
of extreme-mass-ratio mergers, but much work remains before such
simulations are available for generically spinning black holes.

Fortunately our analytical knowledge of the geodesics of the Kerr
metric can help us understand binary black-hole mergers in the
test-particle limit.  These mergers can be broken down into three
stages: inspiral, plunge, and ringdown.  During the inspiral stage,
the orbit of the test particle adiabatically evolves through a series
of geodesics with successively lower energy and angular momentum as
these quantities are radiated away through the emission of
gravitational waves.  This gravitational radiation circularizes
initially eccentric orbits while the evolution is post-Newtonian
\cite{Peters:1963ux}, implying that for many though possibly not all
astrophysical mergers the orbit will have circularized before the test
particle reaches the ISCO.  After this point, during the brief plunge
stage, the test particle rapidly falls into the event horizon of the
larger black hole.  Finally, the quasinormal modes excited during the
merger ring down as the newly formed black hole settles into its final
Kerr configuration.  In the test-particle limit, the energy and
angular momentum radiated during the inspiral stage scales linearly
with $q$, while that radiated during plunge and ringdown scales as a
higher power in the small parameter $q$.  This suggests that one can
reasonably predict the mass and spin of the final black hole by
equating these quantities to the energy and total angular momentum of
the binary at the ISCO.

This prediction assumes that during the inspiral all of the
gravitational waves are radiated outwards to infinity rather than
downwards to the larger black hole's event horizon.  For highly
spinning black holes, these downward gravitational waves will be
superradiantly scattered, extracting energy and angular momentum from
the larger black hole.  Individual modes can be amplified as much as
138\% by a maximally spinning black hole \cite{Teukolsky:1974yv}; the
total energy flux radiated to infinity by a test particle at the ISCO
will be amplified by 12.9\% for a black hole of spin $\chi = 0.999$
\cite{Finn:2000sy}.  Although this amplification decreases rapidly as
a function of the orbital radius, the total energy and angular
momentum extracted throughout the inspiral still scale linearly in $q$
and will therefore remain the dominant factor affecting the final mass
and spin after the energy and angular momentum advected with the test
particle itself during merger.

The primary goal of this paper is to determine how much the
superradiant scattering of gravitational waves during the inspiral
reduces the maximum spin achievable by binary black-hole mergers below
the Kerr limit first predicted by Bardeen \cite{Bardeen:1970}.  In
Sec.~\ref{S:test} we will review the previous analytic predictions for
the final mass and spin that serve as our starting point.  Our method
for calculating the energy and angular momentum extracted from the
spinning black hole will be described in Sec.~\ref{S:super}.  The
results of this calculation are presented in Sec.~\ref{S:res}.  In
Sec.~\ref{S:num} we compare our model to others in the literature,
extrapolate it to comparable-mass mergers and compare with
numerical-relativity simulations, and suggest how it might be combined
with fitting formulas to produce an approximation valid over the
entire domain $0 \leq q \leq 1$.  A brief summary and a few final
remarks on the astrophysical implications of our analysis are offered
in Sec.~\ref{S:disc}.

\section{Test-Particle Mergers} \label{S:test}

Hughes and Blandford \cite{Hughes:2002ei} (hereafter HB) recognized
that in the test-particle limit, the energy and angular momentum
radiated during plunge and ringdown was much less than that released
during the inspiral stage.  They predicted the mass $M_f$ and spin
$S_f$ of the final black hole would therefore be given by
\begin{subequations} \label{E:HB}
  \begin{eqnarray} \label{E:E_HB}
	M_{f, {\rm HB}} &=& m_1 + m_2 E_{\rm ISCO}(\chi_1)~,
	\\ \label{E:L_HB}
	S_{f, {\rm HB}} &=& m_1 m_2 L_{\rm ISCO}(\chi_1) + m_{1}^{2} \chi_1~.
  \end{eqnarray}
\end{subequations}
Here, $E_{\rm ISCO}(\chi)$ is the energy per unit mass of a test
particle on the equatorial ISCO of a Kerr black hole with
dimensionless spin $\chi$.  $L_{\rm ISCO}(\chi)$ is the corresponding
dimensionless orbital angular momentum.  The dimensionless final spin
is simply $\chi_f = S_f/M_{f}^2$.  This formula is exact in the
test-particle limit, and for $q \to 0,~\chi_1
\to 1$ correctly reproduces the Bardeen result
\begin{equation} \label{E:HBlim}
\frac{\partial \chi_{f, {\rm HB}}}{\partial q} \to L_{\rm ISCO}(1)
- 2E_{\rm ISCO}(1) = 0~.
\end{equation}
Maximally spinning black holes cannot be spun up above the Kerr limit.

Although Eqs.~(\ref{E:HB}) are exact in the test-particle limit, they
are manifestly asymmetric in the black-hole labels ``1'' and ``2''.
For example, they ignore altogether the spin $\chi_2$ of the smaller
black hole.  While HB boldly extrapolate their formula to $q = 0.5$,
Buonanno, Kidder, and Lehner \cite{Buonanno:2007sv} (hereafter BKL)
realized that a symmetrized version would more accurately describe
comparable-mass mergers.  They proposed
\begin{subequations} \label{E:BKL}
  \begin{eqnarray} \label{E:E_BKL}
      M_{f, {\rm BKL}} &=& m_1 + m_2~, \\ \label{E:L_BKL}
      S_{f, {\rm BKL}} &=& m_1 m_2 L_{\rm ISCO}(\chi_f) + m_{1}^{2}
      \chi_1 + m_{2}^{2} \chi_2~.
  \end{eqnarray}
\end{subequations}
In addition to the obvious improvement of including the second spin
$\chi_2$, BKL made the inspired choice of using the dimensionless spin
$\chi_f$ of the {\it final} black hole to calculate the orbital
angular momentum of the {\it initial} binary at the ISCO.  Although
counterintuitive at first, this choice correctly captures the total
energy and angular momentum of the system which are assumed to be
conserved after the binary reaches the ISCO.

Equations.~(\ref{E:BKL}) are far more successful at predicting the
final spin from equal-mass mergers than they have any right to be.
They predict that equal-mass nonspinning black holes will merge to
yield a final black hole with spin $\chi_{f, {\rm BKL}} = 0.663$ quite
close to the numerically determined value $\chi_{\rm NR} = 0.68646
\pm 0.00004$ \cite{Scheel:2008rj}.  Equal-mass, maximally spinning
black holes are predicted to produce a final spin $\chi_{f, {\rm BKL}}
= 0.959$ which is also surprisingly close the numerically extrapolated
value $\chi_{\rm NR} = 0.951 \pm 0.004$ \cite{Marronetti:2007wz}.
This latter agreement however is an artifact of overestimating the
final mass in Eq.~(\ref{E:E_BKL}) by failing to account for the energy
radiated during the inspiral.  Overestimating the denominator in the
expression $\chi_f = S_f/M_{f}^{2}$ leads to an underestimate of the
final spin.  In the limit $q \to 0,~\chi_1 \to 1$, this
underestimation implies
\begin{equation} \label{E:BKLlim}
\frac{\partial \chi_{f, {\rm BKL}}}{\partial q} \to L_{\rm ISCO}(1)
- 2 = 2(3^{-1/2} - 1) < 0~.
\end{equation}
Maximally spinning black holes are artificially found to be spun down
by test-particle mergers.  In this model black holes can only be spun
up by test particles to the fictitious limit $\chi_{\rm lim} = 0.948$
at which $\partial \chi_f/\partial q = 0$.

Kesden \cite{Kesden:2008ga} (hereafter K) sought to remedy this by
replacing Eq.~(\ref{E:E_BKL}) with
\begin{equation} \label{E:E_K}
  M_{f, {\rm K}} = M - \mu [1 - E_{\rm ISCO}(\chi_f)]~,
\end{equation}
where $M \equiv m_1 + m_2$ is the sum of the initial masses, $\mu
\equiv m_1m_2/M$ is the reduced mass of the binary, and the energy per
unit mass $E_{\rm ISCO}(\chi_f)$ is evaluated using the final
dimensionless spin in the spirit of BKL.  This formula retains the
desired symmetry of Eqs.~(\ref{E:BKL}) under exchange of black-hole
labels, but also reduces to Eqs.~(\ref{E:HB}) in the test-particle
limit thereby preserving the Bardeen result that black holes can be
spun all the way up to the Kerr limit by test-particle mergers.  It
predicts that equal-mass, nonspinning black holes will merge into a
hole with final spin $\chi_{f, {\rm K,NS}} = 0.687$ in near miraculous
(and probably coincidental) agreement with numerical simulations, but
also predicts an uncomfortably large final spin $\chi_{f, {\rm K,S}} =
0.9988$ for maximally spinning aligned mergers.  Equations
(\ref{E:L_BKL}) and (\ref{E:E_K}) are not the unique choice that
possesses the desired symmetry and limiting behavior.  For example,
replacing Eq.~(\ref{E:L_BKL}) with
\begin{equation} \label{E:L_K}
S_{f, {\rm K}} = \mu M_f L_{\rm ISCO}(\chi_f) + m_{1}^{2} \chi_1 +
      m_{2}^{2} \chi_2
\end{equation}
also maintains these properties, but predicts different results
$\chi_{f, {\rm K,NS}} = 0.675$ and $\chi_{f, {\rm K,S}} = 0.9909$ when
extrapolated to equal-mass mergers.  For our purpose, in this paper of
calculating the maximum final spin for test-particle mergers, either
Eqs.~(\ref{E:E_HB}) or Eq.~(\ref{E:E_K}) paired with either
Eq.~(\ref{E:L_BKL}) or Eq.~(\ref{E:L_K}) can serve as a suitable
starting point.

\section{Superradiant Scattering} \label{S:super}

The gravitational radiation emitted by an inspiraling test particle is
fully described by the complex Weyl scalar
\begin{equation} \label{E:psi4}
\psi_4 \equiv -C_{\alpha\beta\gamma\delta} n^\alpha \bar{m}^\beta
n^\gamma \bar{m}^\delta~,
\end{equation}
where $C_{\alpha\beta\gamma\delta}$ is the Weyl curvature tensor and
$n^\mu$ and $\bar{m}^\mu$ are elements of a Newman-Penrose tetrad of
null 4-vectors \cite{Newman:1961qr}.  Teukolsky
\cite{Teukolsky:1973ha} decomposed $\psi_4$ into multipole moments
\begin{equation} \label{E:psi4MM}
\psi_4 = (r - ia \cos \theta)^{-4} \int_{-\infty}^{\infty} d\omega
\sum_{lm} R_{lm\omega}(r) \,_{-2}S_{lm}^{a\omega}(\theta) e^{im\phi}
e^{-i\omega t}
\end{equation}
and showed that the evolution of the radial modes $R_{lm\omega}(r)$ is
governed by an ordinary second-order differential equation sourced by
the test particle's contribution to the stress-energy tensor.  Sasaki
and Nakamura \cite{Sasaki:1981sx} derived a linear transformation of
$R_{lm\omega}(r)$ that greatly facilitates the solution of this
differential equation.  To solve this equation ourselves, we used the
GREMLIN (Gravitational Radiation in the Extreme Mass ratio LIMit) code
written and generously provided to us by Scott Hughes.  We adopt his
notation and closely follow his treatment presented in
\cite{Hughes:1999bq} to calculate the energy and angular momentum
extracted from the black hole by superradiant scattering during the
inspiral.

The Teukolsky-Sasaki-Nakamura formalism allows one to calculate the
rates $(dE/dt)_{r \to r_+}^{\rm rad}$ and $(dL_z/dt)_{r \to r_+}^{\rm
rad}$ at which energy and angular momentum are radiated down the event
horizon at $r_+$.  The total energy $E_{\rm SR}$ and angular momentum
$J_{\rm SR}$ extracted from the black hole as the test particle
inspirals from infinity to $r$ are integrals of these fluxes
\begin{subequations} \label{E:fSR}
  \begin{eqnarray} \label{E:ESR}
	E_{\rm SR}(\chi, r) &=& \int_{r}^\infty 
	\left( \frac{dE}{dt} \right)_{r \to r_+}^{\rm rad}
	\frac{dr^\prime}{\dot{r}}~, \\ \label{E:JSR}
	J_{\rm SR}(\chi, r) &=& \int_{r}^\infty
	\left( \frac{dL_z}{dt} \right)_{r \to r_+}^{\rm rad}
	\frac{dr^\prime}{\dot{r}}~,
  \end{eqnarray}
\end{subequations}
where the radial velocity $\dot{r}$ during an adiabatic,
quasicircular inspiral is given by
\begin{equation} \label{E:RV}
\dot{r}(\chi, r) = \left( \frac{dE}{dt} \right)_{\rm tot}^{\rm rad}
\left( \frac{dE}{dr} \right)^{-1}~.
\end{equation}
Since $\dot{r}$ is negative as are the fluxes $(dE/dt)_{r \to
r_+}^{\rm rad}$ and $(dL_z/dt)_{r \to r_+}^{\rm rad}$ for high spins,
$E_{\rm SR}$ and $J_{\rm SR}$ are defined to be positive when energy
and angular momentum are extracted from the black hole and negative
when they are absorbed.  Adding the contributions of
Eqs.~(\ref{E:fSR}) to the right-hand sides of Equations (\ref{E:E_K})
and (\ref{E:L_BKL}) yields
\begin{subequations} \label{E:SRp}
  \begin{eqnarray} \label{E:SRpE}
M_f &=& M - \mu [1 - E_{\rm ISCO}(\chi_f)
+ E_{\rm SR}(\chi_f, r_{\rm ISCO})] \quad \quad \\ \label{E:SRpL}
S_f &=& m_1 m_2 [L_{\rm ISCO}(\chi_f) - J_{\rm SR}(\chi_f, r_{\rm ISCO})] \\
&& \quad + m_{1}^{2} \chi_1 + m_{2}^{2} \chi_2 \notag \\ \label{E:SRchi}
\chi_f &=& S_f/M_{f}^{2}
  \end{eqnarray}
\end{subequations}
for our revised prediction for the final black-hole mass and spin.
With a slightly different extrapolation of $S_f$ to large mass ratios,
we can modify Eq.~(\ref{E:L_K}) to obtain
\begin{eqnarray} \label{E:SRpL2}
S_f^\prime &=& \mu M_f [L_{\rm ISCO}(\chi_f)
- J_{\rm SR}(\chi_f, r_{\rm ISCO})] \\
&& \quad + m_{1}^{2} \chi_1 + m_{2}^{2} \chi_2~. \notag
\end{eqnarray}
In the test-particle limit $q \to 0$, Eq.~(\ref{E:SRpE}) combined with
either Eq.~(\ref{E:SRpL}) or Eq.~(\ref{E:SRpL2}) gives
\begin{subequations} \label{E:lim}
  \begin{eqnarray} \label{E:chilim}
\frac{\partial \chi_f}{\partial q}(\chi_1, r) &\equiv&
\frac{\partial \chi_{\rm ISCO}}{\partial q}(\chi_1)
- \frac{\partial \chi_{\rm SR}}{\partial q}(\chi_1, r) \\ \label{E:ISCOlim}
\frac{\partial \chi_{\rm ISCO}}{\partial q}(\chi_1) &\to&
L_{\rm ISCO}(\chi_1) - 2\chi_1 E_{\rm ISCO}(\chi_1) \\ \label{E:SRlim}
\frac{\partial \chi_{\rm SR}}{\partial q}(\chi_1, r) &\to&
J_{\rm SR}(\chi_1, r) - 2\chi_1 E_{\rm SR}(\chi_1, r)~.
  \end{eqnarray}
\end{subequations}
At the limiting spin $\chi_1 = \chi_{\rm lim}$, $\partial
\chi_f(\chi_1, r_{\rm ISCO})/\partial q = 0$, implying that black
holes cannot be spun up beyond $\chi_{\rm lim}$ by test-particle
mergers.

\section{Results} \label{S:res}

\begin{figure}[t!]
\begin{center}
\includegraphics[width=3.5in]{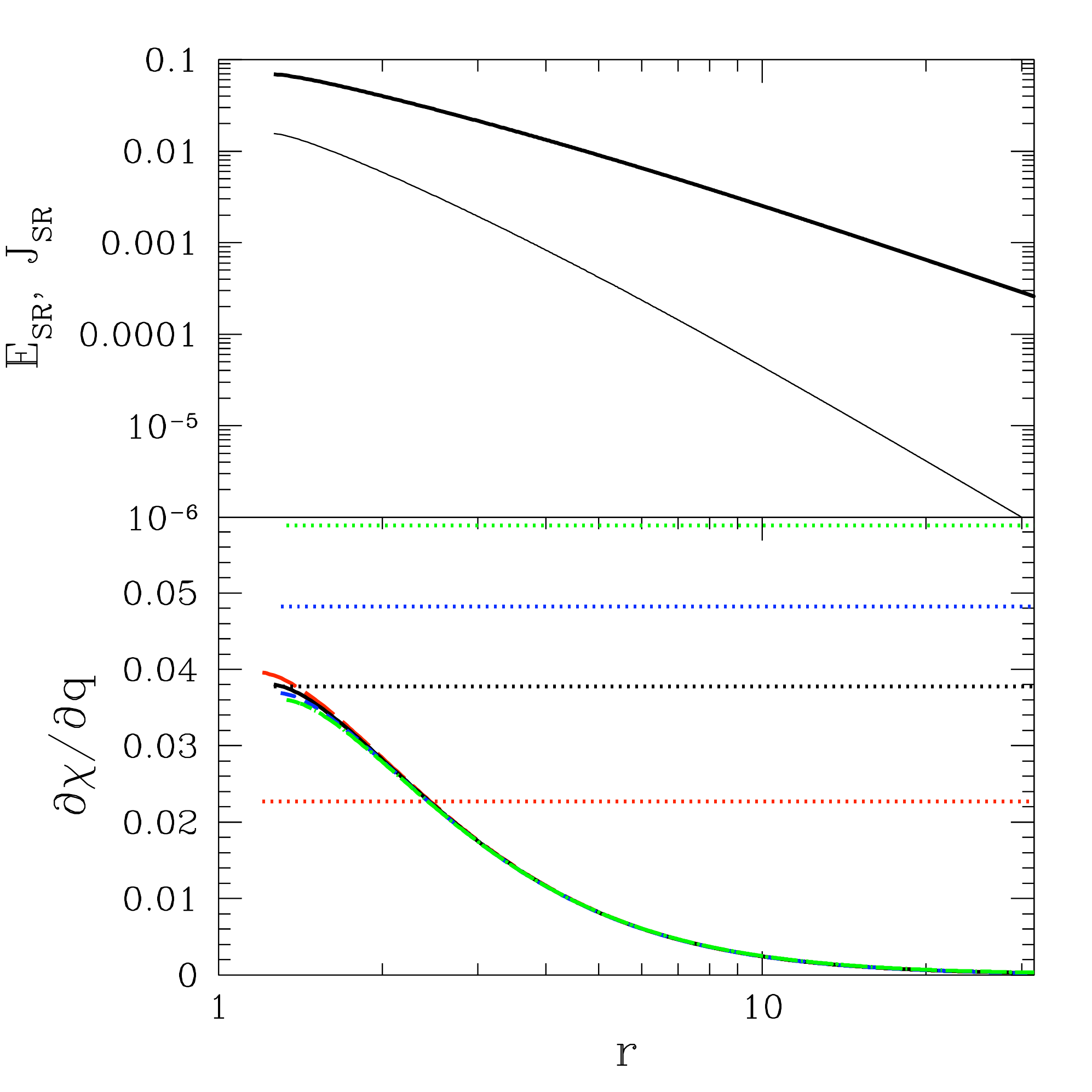}
\end{center}
\caption{{\it Top panel}: The dimensionless energy $E_{\rm SR}$ (thin
line) and angular momentum $J_{\rm SR}$ (thick line) extracted from a
black hole with the limiting spin $\chi_{\rm lim} = 0.9979$ as the
test particle inspirals from infinity to radius $r$. {\it Bottom
panel}: The change in final spin per unit test-particle mass $\partial
\chi/\partial q$.  The dotted horizontal lines show the spin {\it increase}
$\partial \chi_{\rm ISCO}/\partial q$ when the test particle falls
from the ISCO into a black hole of spin $\chi = 0.996$ (green), 0.997
(blue), $\chi_{\rm lim}$ (black), and 0.999 (red).  The curves
$\partial \chi_{\rm SR}/\partial q$ show the spin {\it decrease} as
superradiant scattering extracts energy from black holes with spins of
the corresponding color.  Only at $\chi = \chi_{\rm lim}$ do the lines
and curves intersect at $r_{\rm ISCO}$, indicating that the spin
remains unchanged by the merger.}
\label{F:EJchi}
\end{figure}

In the top panel of Fig.~\ref{F:EJchi}, we show the dimensionless
energy $E_{\rm SR}(\chi, r)$ and angular momentum $J_{\rm SR}(\chi,
r)$ extracted from a black hole of spin $\chi_{\rm lim} = 0.9979$ by
superradiant scattering as a test particle inspirals from infinity to
a radius $r$.  According to Eq.~(\ref{E:SRlim}), these two quantities
determine the spin decrease $\partial \chi_{\rm SR}/\partial q$ due to
the superradiant scattering of downward-going gravitational waves.
This spin decrease is shown for spins $\chi = 0.996$, 0.997,
$\chi_{\rm lim}$, and 0.999 in the bottom panel of Fig.~\ref{F:EJchi}.
The increase in spin $\partial \chi_{\rm ISCO}/\partial q$ when the
test particle itself is accreted by black holes of the same spins is
shown by dotted horizontal lines.  At the limiting spin $\chi_{\rm
lim}$, the total spin decrease $\partial \chi_{\rm SR}/\partial q$
evaluated at the ISCO precisely cancels the spin increase $\partial
\chi_{\rm ISCO}/\partial q$ leaving the dimensionless spin $\chi$
unchanged.  This is seen explicitly by the intersection of the solid
black curve and dotted black line at $r_{\rm ISCO}$ in the bottom
panel of Fig.~\ref{F:EJchi}.  The intersection of $\partial \chi_{\rm
SR}/\partial q$ and $\partial \chi_{\rm ISCO}/\partial q$ at $\chi_1 =
\chi_{\rm lim}$, $r = r_{\rm ISCO}$ implies that the total spin change
$\partial \chi_f/\partial q$ vanishes by Eq.~(\ref{E:chilim}).  Note
that both the spin angular momentum $S_f$ and mass $M_f$ do increase
as a result of the merger, but in just the right ratio as to preserve
$\chi_f = S_f/M_{f}^{2}$. 

The spin increase $\partial \chi_{\rm ISCO}/\partial q$ is a
monotonically decreasing function of $\chi_1$, while the spin decrease
$\partial \chi_{\rm SR}/\partial q$ {\it evaluated at} $r_{\rm ISCO}$
monotonically increases with $\chi_1$.  This implies that the total
spin change $\partial \chi_f/\partial q$ monotonically decreases with
$\chi_1$, intersecting zero at $\chi_1 = \chi_{\rm lim}$.  Black holes
with $\chi_1 < \chi_{\rm lim}$ are spun up by mergers with test
particles on circular equatorial orbits, while black holes with spins
greater than $\chi_{\rm lim}$ are conversely spun down.  Black holes
are {\it never} spun down [$\partial \chi_f(\chi_1, r_{\rm
ISCO})/\partial q \geq 0, \forall \chi_1$] when superradiant
scattering is neglected, reproducing the original Bardeen
\cite{Bardeen:1970} result that the Kerr limit could be saturated.

\begin{figure}[t!]
\begin{center}
\includegraphics[width=3.5in]{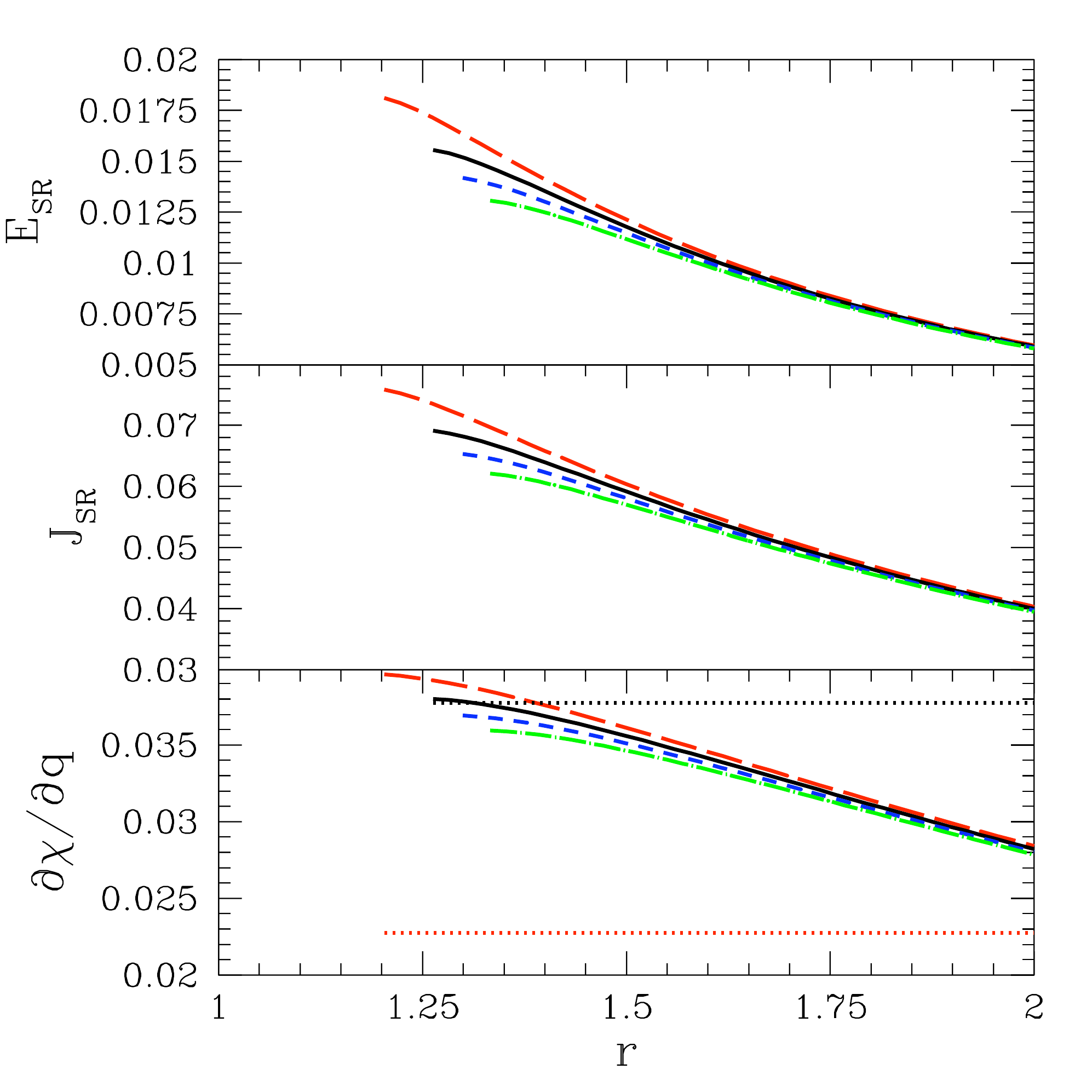}
\end{center}
\caption{The same quantities $E_{\rm SR}$, $J_{\rm SR}$, and $\partial
\chi/\partial q$ presented in Fig.~\ref{F:EJchi} as functions of
the orbital radius $r$.  We zoom into the region near the ISCO to
better display how these quantities evolve near merger.  The colors
and styles of the curves are the same as those in Fig.~\ref{F:EJchi}.}
\label{F:EJchizoom}
\end{figure}

Only one curve for $E_{\rm SR}$ and one for $J_{\rm ISCO}$ were
presented in the top panel of Fig.~\ref{F:EJchi}, as the difference in
these quantities as functions of the initial spin could not be
distinguished in the logarithmic plot needed to depict the five
orders-of magnitude of their evolution.  In Fig.~\ref{F:EJchizoom}, we
zoom in on the region near the ISCO to display this spin dependence.
The total energy and angular momentum extracted are nearly independent
of spin all the way down to $r = 2m_1$, deep within the relativistic
region.  Only very close to the ISCO do the curves diverge, with the
higher values of $E_{\rm SR}$ and $J_{\rm SR}$ for more highly
spinning black holes coming in nearly equal parts from the higher
fluxes at fixed radii and the decreasing value of $r_{\rm ISCO}$ in
the lower bound of the integrals in Eqs.~(\ref{E:fSR}).  

\begin{figure}[t!]
\begin{center}
\includegraphics[width=3.5in]{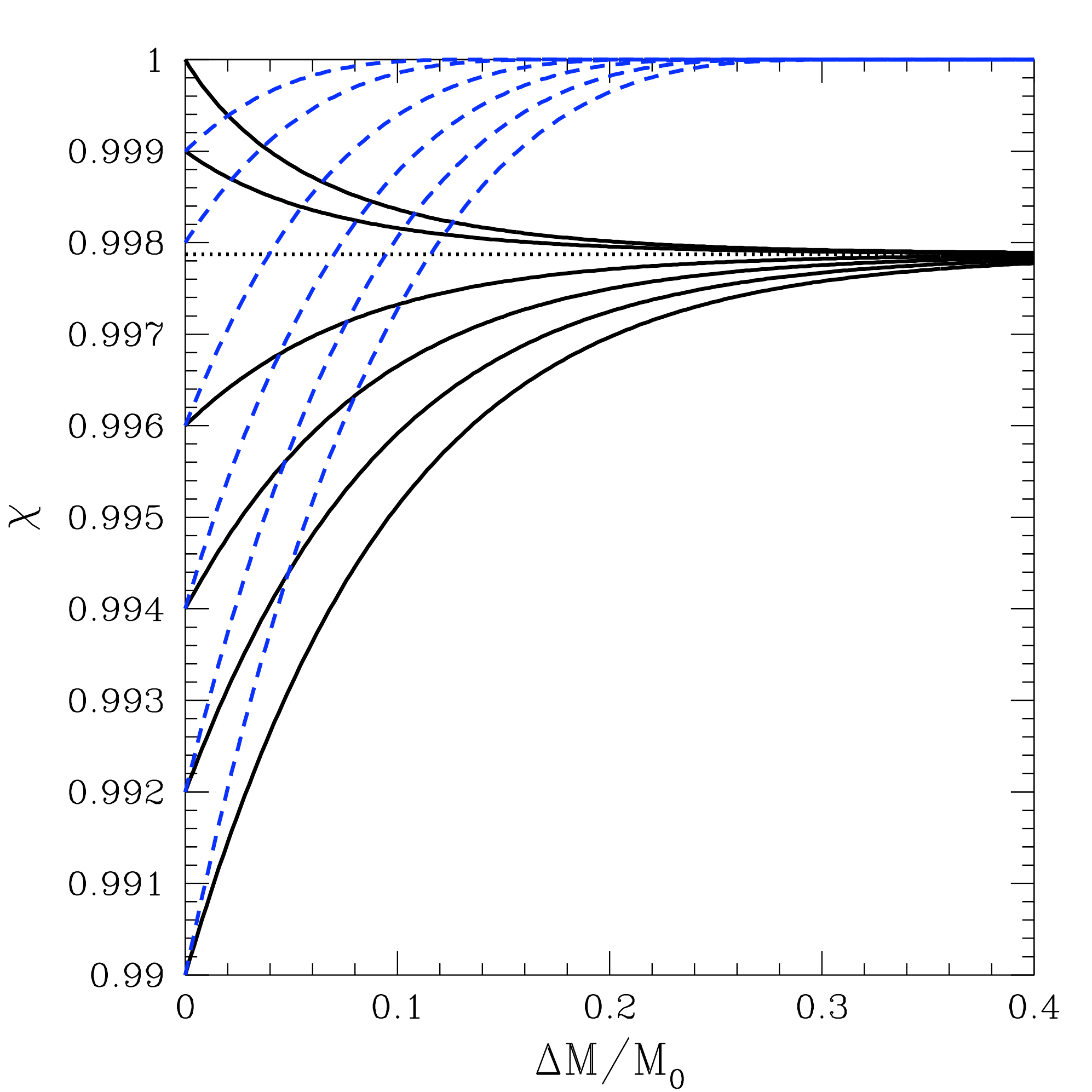}
\end{center}
\caption{The spin $\chi$ of a black hole initially of mass $M_0$
after binary mergers with a mass $\Delta M$ of test particles on
quasicircular equatorial orbits.  The black (solid) curves show the
predictions of our model for different initial spins, while the blue
(dashed) curves show the predicted spins in the absence of scattering.
Spins in our model asymptote to $\chi_{\rm lim} \simeq 0.9979$ as
shown by the horizontal dotted line, while without scattering black
holes can spin all the way up to the Kerr limit.}
\label{F:evochi}
\end{figure}

Using this refined model for test-particle mergers, we can recalculate
how the dimensionless black-hole spin $\chi$ and mass $M$ evolve after
the black hole has accreted a finite mass $\Delta M$ of test particles
gravitationally inspiraling inwards from large radii.  Following
Thorne \cite{Thorne:1974ve}, we find that this evolution is governed
by the coupled first-order differential equations
\begin{subequations} \label{E:evo}
  \begin{eqnarray} \label{E:evochi}
\frac{d\chi}{d\Delta M} &=& \frac{1}{M^2} \frac{dJ}{d\Delta M} -
\frac{2\chi}{M} \frac{dM}{d\Delta M} \\ \label{E:evoM}
\frac{dM}{d\Delta M} &=& E_{\rm ISCO}(\chi)
- E_{\rm SR}(\chi, r_{\rm ISCO}(\chi)) \\ \label{E:evoS}
\frac{dJ}{d\Delta M} &=& M[L_{\rm ISCO}(\chi)
- J_{\rm SR}(\chi, r_{\rm ISCO}(\chi))]~.
  \end{eqnarray}
\end{subequations}
In Fig.~\ref{F:evochi}, we see how the spin $\chi$ evolves for
different values of the initial spin $\chi_0$ (where the curves
intersect the y axis $\Delta M/M_0 = 0$).  The black curves show the
predictions of our model, while the blue curves show those of Bardeen
\cite{Bardeen:1970} without the scattering of gravitational waves by
the black hole.  As expected, in our model black holes with initial
spins $\chi \leq \chi_{\rm lim}$ can only be spun up to this limiting
value no matter how much mass they accrete.  More highly spinning
black holes are spun down to near this limit after accreting $\gtrsim
10\%$ of their initial mass.  By contrast, in the Bardeen model black
holes reach the Kerr limit $\chi = 1$ after accreting a {\it finite}
mass $\Delta M$ of test particles.  For initially nonspinning holes,
this mass is
\begin{equation} \label{E:BarMass}
\Delta M = 3M_0 (\sin^{-1} \sqrt{2/3} - \sin^{-1} 1/3)~.
\end{equation}

\begin{figure}[t!]
\begin{center}
\includegraphics[width=3.5in]{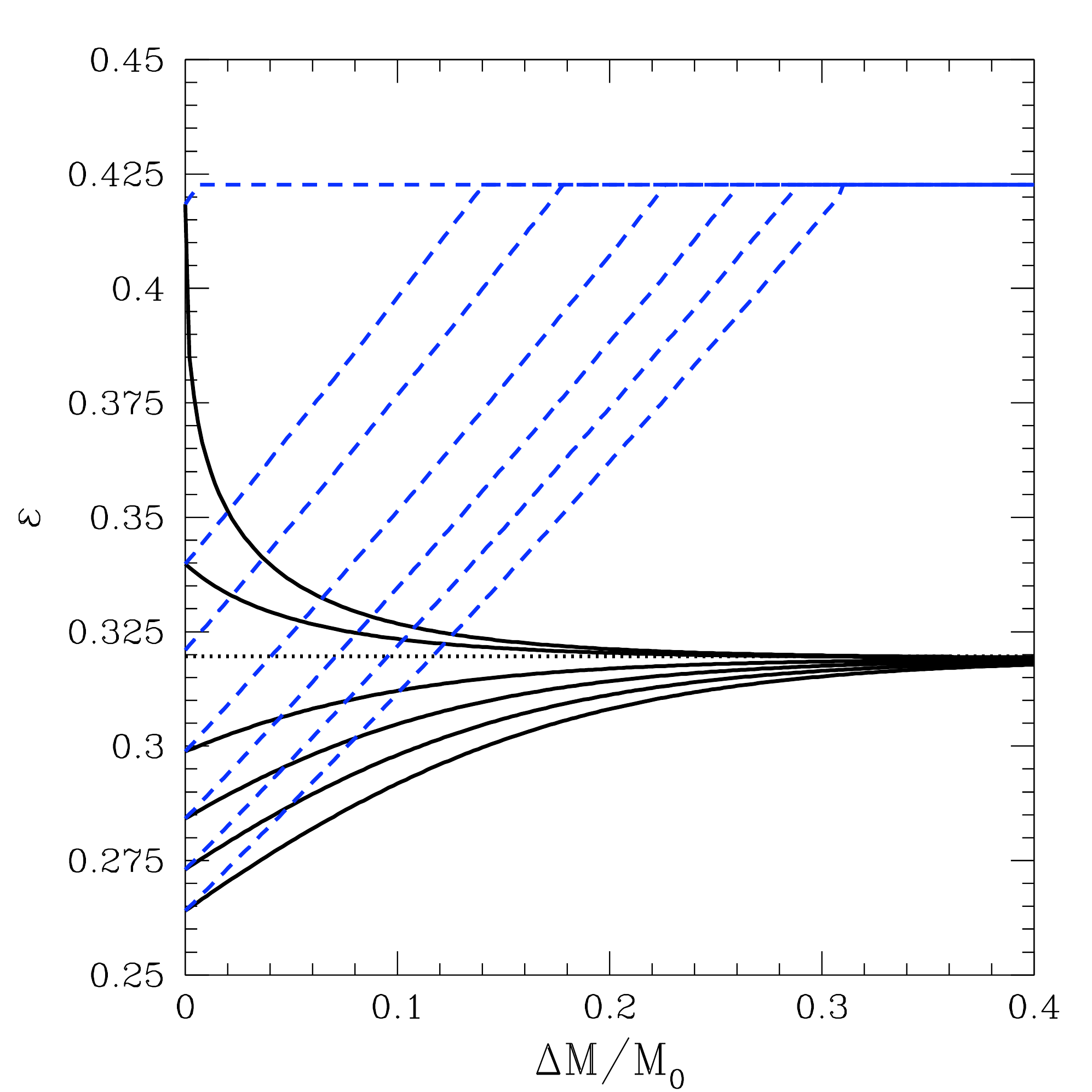}
\end{center}
\caption{The radiative efficiency $\varepsilon = 1 - E_{\rm ISCO}(\chi)$
of the accreting black holes shown in Fig.~\ref{F:evochi}.  In the
Bardeen model (dashed blue curves), the black holes spin up to the
Kerr limit $\chi = 1$ after accreting a finite mass $\Delta M$.  The
radiative efficiency of a maximally spinning black hole is
$\varepsilon = 1 - 1/\sqrt{3} = 0.423$.  In our model (solid black
curves) which includes the superradiant scattering of gravitational
waves, the black hole spins asymptotically approach $\chi_{\rm lim}$.
A black hole with this spin has the substantially lower efficiency
$\varepsilon_{\rm lim} = 0.320$.}
\label{F:eff}
\end{figure}

Radiatively efficient black holes have a luminosity
\begin{equation} \label{E:lum}
L = \varepsilon \frac{d\Delta M}{dt} c^2~,
\end{equation}
where $\varepsilon = 1 - E_{\rm ISCO}(\chi)$ is the radiative
efficiency.  In Fig.~\ref{F:eff}, we show the efficiency $\varepsilon$
of the black holes in Fig.~\ref{F:evochi} as a function of the total
mass $\Delta M$ of test particles they have accreted.  If the
superradiant scattering of gravitational waves is neglected, the black
holes spin up to the Kerr limit $\chi = 1$ after which they are
capable of converting a fraction $\varepsilon = 1 - 1/\sqrt{3} =
0.423$ of accreted mass into radiant energy.  By contrast, accounting
for superradiant scattering reduces the maximum black-hole spin to
$\chi_{\rm lim} = 0.998$.  Although this limiting spin is only 0.2\%
below the Kerr limit, the corresponding limiting radiative efficiency
$\varepsilon_{\rm lim} = 1 - E_{\rm ISCO}(\chi_{\rm lim}) = 0.320$ is
24\% below that of a maximally spinning black hole.  The inability of
astrophysical black to radiate more efficiently than $\chi_{\rm lim}$
will affect efforts to estimate black-hole masses from their
luminosities, such as that by Soltan \cite{Soltan:1982vf} to constrain
the mass function of supermassive black holes from the quasar
luminosity function.

\section{Comparison with Numerical Relativity} \label{S:num}

\begin{figure}[t!]
\begin{center}
\includegraphics[width=3.5in]{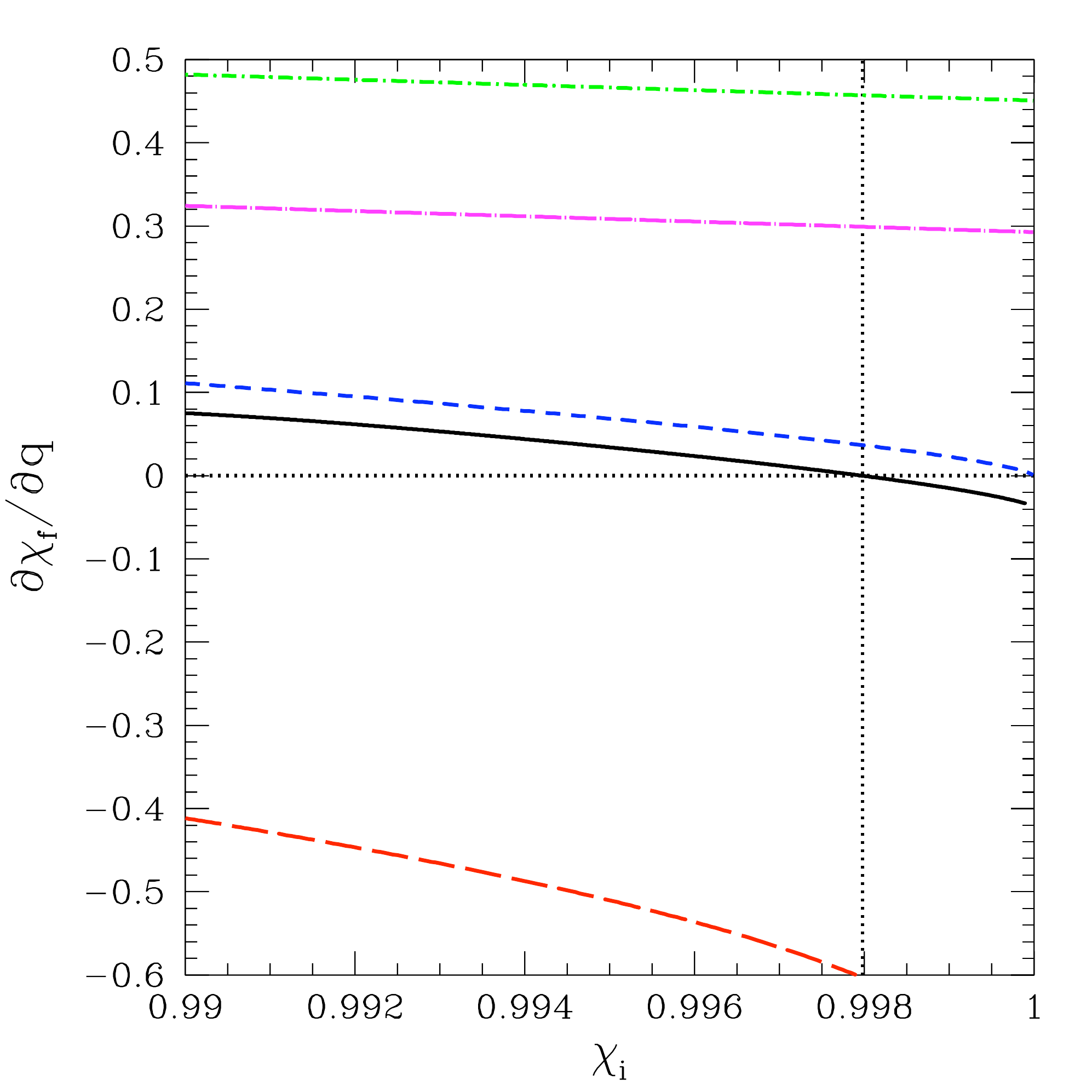}
\end{center}
\caption{The change in final spin per unit test-particle mass
$\partial \chi_f/\partial q$ as a function of the initial spin
$\chi_i$.  The solid black curve shows the predictions of this paper,
with spin down ($\partial \chi_f/\partial q \leq 0$, horizontal dotted
line) possible for spins $\chi_i \geq \chi_{\rm lim}$ (vertical dotted
line).  The short-dashed blue curve shows how this result changes when
superradiant scattering is neglected as in the models of Bardeen
\cite{Bardeen:1970}, HB \cite{Hughes:2002ei}, and K
\cite{Kesden:2008ga}.  The long-dashed red curve gives the BKL
\cite{Buonanno:2007sv} test-particle prediction, while the
short dash-dotted green and long dash-dotted magenta curves are the
predictions of the AEI \cite{Barausse:2009uz} and FAU
\cite{Tichy:2008du} fitting formulas.} \label{F:fitcom}
\end{figure}

How do the predictions of our model compare with those of other
published models?  In Fig.~\ref{F:fitcom}, we again show $\partial
\chi_f/\partial q$, this time as a function of the initial spin
$\chi_i$.  The value of $\chi_i$ for which a given curve crosses the
horizontal dotted line $\partial\chi_f/\partial q = 0$ determines the
maximum spin $\chi_{\rm lim}$ to which a black hole can be spun up by
test-particle mergers.  We see that the superradiant scattering of
gravitational waves produced during the inspiral reduces $\chi_{\rm
lim}$ from the Kerr limit as predicted by Bardeen \cite{Bardeen:1970},
HB \cite{Hughes:2002ei}, and K \cite{Kesden:2008ga} and shown by the
blue (short-dashed) curve to $\chi_i = 0.9979$ shown by the black
(solid) curve.  The red (long-dashed) curve shows the predictions of
the BKL \cite{Buonanno:2007sv} model described in Sec.~\ref{S:test}.
As discussed previously, this model artificially reduces $\chi_{\rm
lim}$ to $\chi_i = 0.948$ by neglecting the spin dependence of the
final mass.  The green (short dash-dotted) and magenta (long
dash-dotted) curves are the predictions of the AEI
\cite{Barausse:2009uz} and FAU \cite{Tichy:2008du} ``fitting
formulas'' for the final spins.

The fitting-formula approach proposes a specific functional form for
the dependence of the given quantity on the parameters $\{ q, \chi_1,
\chi_2 \}$ and a small number of constant coefficients.  These
coefficients are adjusted until the fitting formula best reproduces
the results of a sample of numerical simulations.  Once the fitting
formula has been calibrated in this manner, it should be able to
predict the results of any future simulations.  The AEI and FAU
fitting formulas were calibrated with an extensive sample of numerical
simulations of binary mergers with varying mass ratios and spins, none
of which produced a black hole whose final spin exceeded the Kerr
limit.  Yet Fig.~\ref{F:fitcom} shows that both curves have
$\partial\chi_f/\partial q > 0$ even for $\chi_i = 1$, implying that
test-particle mergers can spin black holes above the Kerr limit.  One
possible explanation of this unphysical prediction is that the fitted
coefficients were misdetermined because of the limited range of mass
ratios in the sample of simulations with which they were calibrated.
A second explanation is that the proposed functional forms are
inadequate to describe the mass-ratio or spin dependence of the final
spin for {\it any} choice of coefficients.  Further numerical
simulations, particularly at smaller mass ratios, are needed to
determine which of these explanations is correct.  This problem is not
just restricted to test-particle mergers; the AEI and FAU formulas
predict final spins above the Kerr limit for mass ratios as large as
$q = 0.283$ and $q = 0.2434$, respectively.

Although the AEI and FAU fitting formulas break down in the
test-particle limit, by design they agree closely with simulations of
comparable-mass mergers like those with which they were calibrated.
In Table~\ref{T:NRcomp} we compare our model, extrapolated to
equal-mass mergers according to Eqs.~(\ref{E:SRp}), with both the
fitting formulas and the most accurate numerical simulations of
equal-mass ($q = 1$), equal-spin ($\chi_1 = \chi_2 = \chi_i$), aligned
mergers.  Both the nonspinning ($\chi_i = 0$) \cite{Scheel:2008rj}
and antialigned ($\chi_i = -0.43757$) \cite{Chu:2009md} simulations
were performed with the spectral-methods code developed by the
Caltech-Cornell group.  The $\chi_i = 1$ numerical result
\cite{Marronetti:2007wz} listed in the third column is an
extrapolation to maximal spins of a series of simulations produced
with the BAM finite-difference code used by the Jena and FAU groups.

\begin{table}
  \begin{footnotesize}
    \begin{center}
      \begin{tabular}{|c||c|c|c|}
	\hline
	& \multicolumn{3}{c |}
	{Initial spins $\chi_i$} \\
        \hline
        $\chi_i$ & -0.43757 \cite{Chu:2009md} &
	0 \cite{Scheel:2008rj} & 1 \cite{Marronetti:2007wz} \\
        \hline \hline
	& \multicolumn{3}{c |}{Final spins $\chi_f$} \\
	\hline
        NR & $0.547812 \pm 0.000009$ & 
	$0.68646 \pm 0.00004$ & $0.951 \pm 0.004$ \\
        \hline
	KLP & 0.520861 & 0.686354 & 0.996439 \\
        \hline
	KLP$^\prime$ & 0.509269 & 0.674197 & 0.986947 \\
	\hline
	K \cite{Kesden:2008ga} & 0.521153 & 0.687036 & 0.998805 \\
	\hline
	BKL \cite{Buonanno:2007sv} & 0.505148 & 0.663086 & 0.959107 \\
	\hline
	AEI \cite{Barausse:2009uz} & 0.546646 & 0.686460 & 0.961491 \\
	\hline
	FAU \cite{Tichy:2008du} & 0.548602 & 0.6860 & 0.9540 \\
	\hline
	BK \cite{Boyle:2007ru} & 0.547562 & 0.6893 & 0.9504 \\
	\hline \hline
	& \multicolumn{3}{c |}{Final masses $M_f/M$} \\
	\hline
	NR & $0.961109 \pm 0.000003$ & 
	$0.95162 \pm 0.00002$ & $...$ \\
	\hline
	KLP & 0.979028 & 0.974530 & 0.920918 \\
        \hline
	KLP$^\prime$ & 0.979268 & 0.974984 & 0.932995 \\
	\hline
	K \cite{Kesden:2008ga} & 0.979039 & 0.974565 & 0.916181 \\
        \hline
	FAU \cite{Tichy:2008du} & 0.962877 & 0.9515 & 0.9255 \\
	\hline
	BK \cite{Boyle:2007ru} & 0.964034 & 0.9530 & 0.9009 \\
	\hline
      \end{tabular}
    \end{center}
  \end{footnotesize}
\caption{A comparison between the final spins and masses determined
	by numerical simulations and those predicted by various
	fitting formulas.  All three simulations begin with equal-mass
	binaries whose spins are aligned or antialigned with the
	orbital angular momentum and have magnitudes given in the
	first row.  These simulations are described more fully in the
	references provided.  The next 8 rows give the final spin
	$\chi_f$ as determined by NR or predicted in the referenced
	papers.  KLP are the predictions of this paper.  The final 6 rows
	give the corresponding predictions for the final mass $M_f/M$.}
	\label{T:NRcomp}
\end{table}

The first row gives the calculated values of the final spin $\chi_f$
for these three numerical-relativity (NR) simulations.  The next 7
rows give the final spins predicted for these binaries by many
different published formulas.  The second row lists the predictions of
this paper, Kesden-Lockhart-Phinney (KLP).  Our prediction for the
nonspinning merger agrees with this numerical result to about
$10^{-4}$.  An agreement this good between a {\it test-particle}
extrapolation and an {\it equal-mass} simulation can only be a
coincidence, as is demonstrated by the third row KLP$^\prime$.  Here
we have substituted $S_{f}^{\prime}$ from Eq.~(\ref{E:SRpL2}) for
$S_f$ from Eq.~(\ref{E:SRpL}) in our prediction $\chi_f =
S_f/M_{f}^{2}$.  Although $S_{f}^{\prime}$ has the same symmetries and
limiting behavior as $S_f$, the prediction changes by about 1\% when
extrapolating all the way to equal masses.  Agreement beyond this
accuracy must be considered coincidental unless we discover a
fundamental reason to prefer $S_f$ over $S_{f}^{\prime}$.

The fourth row lists the predictions of Kesden \cite{Kesden:2008ga},
which was the starting point for this paper but neglected the
superradiant scattering of downward-going radiation.  Comparing the
rows labeled KLP and K, we see that this scattering only reduces the
final spin by about 0.25\% even for binaries initially spinning at the
Kerr limit.  Although this effect seems negligible, it is potentially
detectable since several gravitational-wave observables depend very
sensitively on $\chi$ near the Kerr limit.  We shall elaborate on this
in the final discussion in Sec~\ref{S:disc}.  Rows 5 through 8 list
the predictions of other published formulas.  We have already
discussed the BKL model adequately; its close agreement with the
extrapolation to maximally spinning binaries is a fortuitous
coincidence.  The AEI, FAU, and BK \cite{Boyle:2007ru} fitting
formulas all do excellent jobs of reproducing the numerical results,
though the exact agreement of the AEI formula with the nonspinning
simulation results from this simulation being included in the set with
which this formula was calibrated.

The final 6 rows of Table~\ref{T:NRcomp} show how our predicted final
masses $M_f$, extrapolated from the test-particle limit as before,
compare with the numerical simulations and numerically calibrated
fitting formulas.  Our predictions overestimate $M_f$ compared to the
simulations and fitting-formula predictions for the antialigned and
nonspinning mergers.  This may result from our failure to account for
the energy carried away by gravitational-wave emission {\it after} the
merger.  While this emission becomes negligible compared to that during
the inspiral as $q \to 0$, significant emission after the formation of
a common horizon can occur for equal-mass mergers.  We plan to
incorporate radiation during the plunge and ringdown stages of the
merger in future work.  The final mass has not been reliably
determined for the maximally spinning case, which perhaps accounts for
the greater discrepancy between the different fitting formulas.  These
formulas included only a limited number of highly spinning mergers in
the set with which they were calibrated.

It is interesting to note when comparing rows K and KLP that
superradiant scattering actually {\it increases} the predicted final
mass for the maximally spinning merger, unlike in the other two cases.
This is counterintuitive, since the top panel of
Fig.~\ref{F:EJchizoom} indicates that superradiant scattering extracts
the {\it most} energy $E_{\rm SR}(\chi, r_{\rm ISCO})$ from the most
highly spinning black holes.  However, as the test particle inspirals,
superradiant scattering reduces the spin $\chi$, which moves the ISCO
radius $r_{\rm ISCO}(\chi)$ outwards and increases the ISCO energy
$E_{\rm ISCO}(\chi)$.  For large spins, the steep increase in $E_{\rm
ISCO}(\chi)$ with spin more than compensates for the additional energy
$E_{\rm SR}(\chi, r_{\rm ISCO})$ extracted, thus increasing the final
mass $M_f$.

Until numerical relativists can accurately simulate mergers with high
spins and small mass ratios in an acceptable amount of time, we will
need to rely on a combination of fitting formulas and test-particle
extrapolations.  These two methods complement each other, and any
model that attempts to make predictions throughout the entire region
$0 \leq q \leq 1$ should take advantage of both approaches.  The
fitting formulas are most accurate predicting the result of mergers
close in parameter space to the simulations with which they were
calibrated.  Currently this consists mostly of comparable-mass mergers
with $q \geq 0.1$.  The test-particle extrapolations like the one
proposed in this paper are most reliable for $q \ll 1$.  We can
readily modify the AEI formula \cite{Barausse:2009uz} for aligned
mergers
\begin{equation} \label{E:AEI}
\chi_f = \tilde{\chi} + \tilde{\chi}\eta(s_4\tilde{\chi} + s_5\eta + t_0)
+ \eta(2\sqrt{3} + t_2\eta + t_3\eta^2)~,
\end{equation}
with
\begin{equation} \label{E:atil}
\tilde{\chi} \equiv \frac{\chi_1 + q^2 \chi_2}{1 + q^2}~,
\end{equation}
to incorporate the results of this paper.  The first step in this
modification was already taken in Rezzolla {\it et al.}
\cite{Rezzolla:2007rd}, where it was recognized that setting the
coefficient of the term proportional to $\eta$ equal to $2\sqrt{3}$
would reproduce the test-particle prediction for nonspinning black
holes in the absence of superradiant scattering of gravitational
waves.  We propose that {\it all} terms linearly proportional to
$\eta$ can be replaced by our result for $\partial \chi_f/\partial q$
from Sec.~\ref{S:res}
\begin{equation} \label{E:AEIimp}
\chi_{\rm KLP} = \tilde{\chi} +
\frac{\partial \chi_f}{\partial q}(\tilde{\chi})\eta +
(t_2 + s_5\tilde{\chi})\eta^2 + t_3\eta^3~.
\end{equation}
This eliminates two of the coefficients from Eq.~(\ref{E:AEI}) and
guarantees that the formula behaves properly in the test-particle
limit.  The highly accurate equal-mass nonspinning simulation
\cite{Scheel:2008rj} can still be used to determine $t_2$ as in
Barausse and Rezzolla \cite{Barausse:2009uz} if desired.  If the
formula no longer has enough degrees of freedom to reproduce existing
numerical results, additional terms proportional to highers powers of
$\eta$ and $\tilde{\chi}$ can be added.  These new terms will not
affect the test-particle behavior, as $\eta \to 0$ in this limit, but
can help to fit comparable-mass mergers where $\eta \lesssim 0.25$.
There also remains additional freedom in our choice of $\tilde{\chi}$.
Although the choice given in Eq.~(\ref{E:atil}) possesses the desired
symmetry and limiting behavior
\begin{subequations} \label{E:qlim}
  \begin{eqnarray} \label{E:qlim0}
	q \to 0 &\Rightarrow& \tilde{\chi} \to \chi_1 \\ \label{E:qlimI}
	q \to \infty &\Rightarrow& \tilde{\chi} \to \chi_2~,
  \end{eqnarray}
\end{subequations}
this choice is not unique and alternatives may prove more suitable.
An iterative approach for determining $\tilde{\chi}$ as a function of
$\{ q, \chi_1, \chi_2 \}$ as in BKL \cite{Buonanno:2007sv} and Kesden
\cite{Kesden:2008ga} should be explored as well.

This same approach can be used to improve fitting formulas for the
final mass.  Tichy and Marronetti \cite{Tichy:2008du} proposed the
fitting formula
\begin{equation} \label{E:FAU}
\frac{M_f}{M} = 1 + 4(m^0 - 1)\eta + 16m^{a1}\eta^2(\chi_1 + \chi_2)~,
\end{equation}
where $m^0$ and $m^{a1}$ are fitting coefficients.  In the
test-particle limit $\eta \to q \to 0$ this reduces to
\begin{equation} \label{E:FAUtest}
\frac{\partial}{\partial\eta} \left( \frac{M_f}{M} \right) = 4(m^0 - 1) = -0.194~.
\end{equation}
Our Eq.~(\ref{E:SRpE}) in this limit predicts
\begin{equation} \label{E:testMf}
\frac{\partial}{\partial \eta} \left( \frac{M_{\rm KLP}}{M} \right) =
E_{\rm ISCO}(\chi_1) - E_{\rm SR}(\chi_1, r_{\rm ISCO}) - 1,
\end{equation}
correctly capturing the spin dependence of the binding energy $E_{\rm
ISCO}$ and extracted energy $E_{\rm SR}$ on the larger black hole's
spin.  We can use this result to replace the term linear in $\eta$ in
Eq.~(\ref{E:FAU}) to yield
\begin{equation} \label{E:FAUimp}
\frac{M_{\rm KLP}}{M} = 1 + \frac{\partial}{\partial \eta}
\left(\frac{M_{\rm KLP}}{M}\right)
(\tilde{\chi})\eta + 32m^{a1}\eta^2\tilde{\chi}~.
\end{equation}
Notice that we have also replaced the sum of the spins in the third
term with $\tilde{\chi}$, as the sum did not possess the required
limiting behavior of Eq.~(\ref{E:qlim}).  Additional terms
proportional to higher powers in $\eta$ can be added to improve the
agreement with comparable-mass simulations {\it without} affecting the
test-particle behavior.

\section{Discussion} \label{S:disc}

The superradiant scattering of gravitational waves emitted during the
inspiral sets a fundamental upper limit $\chi_{\rm lim} = 0.9979 \pm
0.0001$ on the spin a black hole may attain by accreting test
particles on quasicircular equatorial orbits.  For this limiting
spin, the energy and angular momentum advected when the test particle
falls from the ISCO combines with that extracted by superradiant
scattering to leave the spin $\chi = S/M^2$ of the black hole
unchanged.  Black-hole spins greater than $\chi_{\rm lim}$ will be
reduced by mergers with test particles on quasicircular equatorial
orbits, even though the physical angular momentum $S$ and mass $M$
will both increase.

Jacobson and Sotiriou \cite{Jacobson:2009kt} recently argued that
small but finite-mass particles on highly {\it hyperbolic} orbits
could spin black holes above $\chi_{\lim}$ and even above the Kerr
limit itself.  However, this study neglected gravitational radiation
and any superradiant enhancement, which could be considerable for
these highly spinning mergers.  Such hyperbolic orbits are also
extremely unlikely in any realistic astrophysical scenario.

Our limit is nearly indistinguishable from the limit $\chi_{\rm gas} =
0.9980 \pm 0.0002$ for black holes grown through thin-disk gas
accretion \cite{Thorne:1974ve}, suggesting that spin measurements
alone cannot determine the relative contribution of these two channels
of black-hole growth.  If gas accretion occurs through an
advection-dominated accretion flow (ADAF) rather than a thin disk, the
limiting black-hole spin is reduced to $\chi_{\rm ADAF} \simeq 0.96$,
and the launching of jets can further reduce this spin to $\chi_{\rm
jets} \simeq 0.93$ \cite{Benson:2009kx}.  Since our limit $\chi_{\rm
lim}$ is larger than for other modes of black-hole growth, it seems to
be a fairly robust upper bound on spins that can be obtained by {\it
any} astrophysical means.

Although SBHs are grown primarily through gas accretion rather than
test-particle mergers, our limiting spin $\chi_{\rm lim}$ can still be
astrophysically relevant.  Gas accretion occurs episodically, when
galactic major mergers produce global torques that funnel large
amounts of gas to galactic centers \cite{Mihos:1995ri,Barnes:1996qt}.
Accretion of this gas onto the SBH leads to AGN feedback
\cite{Silk:1997xw} which simulations demonstrate is capable of
suppressing further accretion and black-hole growth
\cite{DiMatteo:2005sd}.  By contrast, compact objects such as white
dwarfs, neutron stars, and stellar-mass black holes (effectively test
particles when orbiting an SBH) will inspiral continuously as they are
scattered into the ``loss cone'' of orbits for which gravitational
inspiral occurs more rapidly than subsequent scattering
\cite{Frank:1976uy}.  Although the same global torques driven by
galactic major mergers may enhance the rate of these
extreme-mass-ratio inspirals (EMRIs), other dynamical processes will
refill the loss cone even between periods of AGN activity
\cite{Sigurdsson:2003ei}.  Event rates for EMRIs depend sensitively on
highly uncertain factors such as the dynamical state, star-formation
history, binary fraction, and initial mass function within nuclear
star clusters.  The capture rate for $10 M_\odot$ black holes by a
$10^6 M_\odot$ SBH can be as high as $10^{-4} {\rm yr}^{-1}$
immediately following a nuclear starburst, and could average $10^{-6}
{\rm yr}^{-1}$ for nucleated spiral galaxies like the Milky Way
\cite{Sigurdsson:1996uz}.  If such a galaxy has not had a major
merger since redshift $z \simeq 1$ (about 7.5 Gyr ago), its SBH would
aqcuire a fraction
\begin{equation} \label{E:EMRIf}
\frac{\Delta M}{M_0} \simeq \frac{10 M_\odot \times 10^{-6}{\rm yr}^{-1}
\times 7.5~{\rm Gyr}}{10^6 M_\odot} = 0.075
\end{equation}
of its mass through EMRIs.  We see from Fig.~\ref{F:evochi} that this
is a large enough fraction to drive the SBH spin towards our limiting
value $\chi_{\rm lim}$ from the initial spins $\chi_0 \gtrsim 0.9$
expected from gas accretion.  This suggests that EMRIs can potentially
affect SBH spin evolution, though modest gas accretion can also occur
between major mergers and not all EMRIs will be on circular,
equatorial orbits.

Even if our limiting spin $\chi_{\rm lim}$ can be realized
astrophysically, does its marginal difference from the Kerr limit have
any observable consequences?  The binding energy per unit mass at the
ISCO $1 - E_{\rm ISCO}$ depends very sensitively on $\chi$ near the
Kerr limit, and sets an upper bound on the AGN efficiency $\varepsilon
\equiv L/\dot{M}c^2$, where $L$ is the AGN luminosity, $\dot{M}$ is
the accretion rate, and $c$ is the speed of light.  Soltan
\cite{Soltan:1982vf} argued that the total cosmological mass in SBHs
could be estimated by equating the observed energy emitted by AGN to
that released during theoretical models of SBH growth.  If spins at
the Kerr limit were used for an analysis of this kind instead of our
limiting spin $\chi_{\rm lim} = 0.9979$, the AGN efficiency
$\varepsilon$ would be overestimated by $E_{\rm ISCO}(\chi_{\rm lim})
- E_{\rm ISCO}(1) = 0.1031$ leading to a corresponding underestimate
in the total SBH mass.

In addition to their influence on the SBH spin, the EMRIs themselves
are an important source for gravitational-wave detectors such as LIGO
and LISA.  Decreasing the spin from the Kerr limit to $\chi_{\rm lim}$
moves the ISCO radius out from $r_{\rm ISCO}(1) = m_1$ to $r_{\rm
ISCO}(\chi_{\rm lim}) = 1.242 m_1$, and decreases the ISCO orbital
frequency from $\Omega_{\rm ISCO}(1) = 0.50 m_{1}^{-1}$ to
$\Omega_{\rm ISCO}(\chi_{\rm lim}) = 0.42 m_{1}^{-1}$.  This decrease
could be observable in high signal-to-noise EMRIs seen by both these
experiments.  Our limiting spin $\chi_{\rm lim}$ is thus not only an
interesting consequence of general relativity, but also one that has
potentially observable implications for astrophysics and
gravitational-wave detection.
	
\vspace{0.3cm}

{\bf Acknowledgements.}  We thank Scott Hughes for generously allowing
us to use a version of his GREMLIN (Gravitational Radiation in the
Extreme Mass ratio LIMit) code to calculate the energy and angular
momentum radiated during the inspiral.  We also thank Andrew Benson,
Chris Hirata, Pedro Marronetti, Uli Sperhake, and Kip Thorne for
useful conversations.  The authors acknowledge support from NASA BEFS
Grant No.~NNX07AH06G (PI:Phinney).  M. Kesden also acknowledges support
from NSF Grant No.~PHY-0601459 (PI: Thorne).


\begin{thebibliography}{99}

\bibitem{Kerr:1963ud}
  R.~P.~Kerr,
  Phys.\ Rev.\ Lett.\  {\bf 11}, 237 (1963).

\bibitem{Reynolds:1998ie}
  C.~S.~Reynolds, A.~J.~Young, M.~C.~Begelman and A.~C.~Fabian,
  Astrophys.\ J.\ {\bf 514}, 164 (1999).

\bibitem{Brenneman:2006hw}
  L.~W.~Brenneman and C.~S.~Reynolds,
  Astrophys.\ J.\  {\bf 652}, 1028 (2006).

\bibitem{Bardeen:1970}
  J.~M.~Bardeen,
  Nature\  {\bf 226}, 64 (1970).

\bibitem{Shakura:1972te}
  N.~I.~Shakura and R.~A.~Sunyaev,
  Astron.\ Astrophys.\  {\bf 24}, 337 (1973).

\bibitem{Godfrey:1970am}
  B.~B.~Godfrey,
  Phys.\ Rev.\  D {\bf 1}, 2721 (1970).

\bibitem{Thorne:1974ve}
  K.~S.~Thorne,
  Astrophys.\ J.\  {\bf 191}, 507 (1974).

\bibitem{Popham:1998iw}
  R.~Popham and C.~F.~Gammie,
  Astrophys.\ J.\  {\bf 504}, 419 (1998)
  [arXiv:astro-ph/9802321].

\bibitem{Balbus:1991ay}
  S.~A.~Balbus and J.~F.~Hawley,
  Astrophys.\ J.\  {\bf 376}, 214 (1991).

\bibitem{Gammie:2003qi}
  C.~F.~Gammie, S.~L.~Shapiro and J.~C.~McKinney,
  Astrophys.\ J.\  {\bf 602}, 312 (2004)
  [arXiv:astro-ph/0310886].

\bibitem{Benson:2009kx}
  A.~J.~Benson and A.~Babul,
  Mon.\ Not.\ Roy.\ Astron.\ Soc.\  {\bf 397}, 1302 (2009)
  [arXiv:0905.2378 [astro-ph.HE]].

\bibitem{Scheel:2008rj}
  M.~A.~Scheel, M.~Boyle, T.~Chu, L.~E.~Kidder, K.~D.~Matthews and H.~P.~Pfeiffer,
  Phys.\ Rev.\  D {\bf 79}, 024003 (2009)
  [arXiv:0810.1767 [gr-qc]].

\bibitem{Chu:2009md}
  T.~Chu, H.~P.~Pfeiffer and M.~A.~Scheel,
  arXiv:0909.1313 [gr-qc].

\bibitem{Lovelace:2008tw}
  G.~Lovelace, R.~Owen, H.~P.~Pfeiffer and T.~Chu,
  Phys.\ Rev.\  D {\bf 78}, 084017 (2008)
  [arXiv:0805.4192 [gr-qc]].

\bibitem{Marronetti:2007wz}
  P.~Marronetti, W.~Tichy, B.~Brugmann, J.~Gonzalez and U.~Sperhake,
  Phys.\ Rev.\  D {\bf 77}, 064010 (2008).

\bibitem{Lousto:2010ut}
  C.~O.~Lousto, Y.~Zlochower,
  [arXiv:1009.0292 [gr-qc]].

\bibitem{Gonzalez:2008bi}
  J.~A.~Gonzalez, U.~Sperhake and B.~Brugmann,
  Phys.\ Rev.\  D {\bf 79}, 124006 (2009)
  [arXiv:0811.3952 [gr-qc]].

\bibitem{Lousto:2010tb}
  C.~O.~Lousto, H.~Nakano, Y.~Zlochower {\it et al.},
  Phys.\ Rev.\ Lett.\  {\bf 104}, 211101 (2010).
  [arXiv:1001.2316 [gr-qc]].

\bibitem{Lousto:2010qx}
  C.~O.~Lousto, H.~Nakano, Y.~Zlochower {\it et al.},
  [arXiv:1008.4360 [gr-qc]].

\bibitem{Peters:1963ux}
  P.~C.~Peters and J.~Mathews,
  Phys.\ Rev.\  {\bf 131}, 435 (1963).

\bibitem{Teukolsky:1974yv}
  S.~A.~Teukolsky and W.~H.~Press,
  Astrophys.\ J.\  {\bf 193}, 443 (1974).

\bibitem{Finn:2000sy}
  L.~S.~Finn and K.~S.~Thorne,
  Phys.\ Rev.\  D {\bf 62}, 124021 (2000)
  [arXiv:gr-qc/0007074].

\bibitem{Hughes:2002ei}
  S.~A.~Hughes and R.~D.~Blandford,
  Astrophys.\ J.\  {\bf 585}, L101 (2003).

\bibitem{Buonanno:2007sv}
  A.~Buonanno, L.~E.~Kidder and L.~Lehner,
  Phys.\ Rev.\  D {\bf 77}, 026004 (2008).

\bibitem{Kesden:2008ga}
  M.~Kesden,
  Phys.\ Rev.\  D {\bf 78}, 084030 (2008)
  [arXiv:0807.3043 [astro-ph]].

\bibitem{Newman:1961qr}
  E.~Newman and R.~Penrose,
  J.\ Math.\ Phys.\  {\bf 3}, 566 (1962).

\bibitem{Teukolsky:1973ha}
  S.~A.~Teukolsky,
  Astrophys.\ J.\  {\bf 185}, 635 (1973).

\bibitem{Sasaki:1981sx}
  M.~Sasaki and T.~Nakamura,
  Prog.\ Theor.\ Phys.\  {\bf 67}, 1788 (1982).

\bibitem{Hughes:1999bq}
  S.~A.~Hughes,
  Phys.\ Rev.\  D {\bf 61}, 084004 (2000)
  [Erratum-ibid.\  D {\bf 63}, 049902 (2001\ ERRAT,D65,069902.2002\ ERRAT,D67,089901.2003)]
  [arXiv:gr-qc/9910091].

\bibitem{Soltan:1982vf}
  A.~Soltan,
  Mon.\ Not.\ Roy.\ Astron.\ Soc.\  {\bf 200}, 114 (1982).

\bibitem{Barausse:2009uz}
  E.~Barausse and L.~Rezzolla,
  arXiv:0904.2577 [gr-qc].

\bibitem{Tichy:2008du}
  W.~Tichy and P.~Marronetti,
  Phys.\ Rev.\  D {\bf 78}, 081501 (2008)
  [arXiv:0807.2985 [gr-qc]].

\bibitem{Boyle:2007ru}
  L.~Boyle and M.~Kesden,
  Phys.\ Rev.\  D {\bf 78}, 024017 (2008)
  [arXiv:0712.2819 [astro-ph]].

\bibitem{Rezzolla:2007rd}
  L.~Rezzolla, P.~Diener, E.~N.~Dorband, D.~Pollney, C.~Reisswig, E.~Schnetter and J.~Seiler,
  Astrophys.\ J.\  {\bf 674}, L29 (2008)
  [arXiv:0710.3345 [gr-qc]].

\bibitem{Jacobson:2009kt}
  T.~Jacobson, T.~P.~Sotiriou,
  Phys.\ Rev.\ Lett.\  {\bf 103}, 141101 (2009).
  [arXiv:0907.4146 [gr-qc]].

\bibitem{Mihos:1995ri}
  J.~C.~Mihos and L.~Hernquist,
  Astrophys.\ J.\  {\bf 464}, 641 (1996)
  [arXiv:astro-ph/9512099].

\bibitem{Barnes:1996qt}
  J.~E.~Barnes and L.~Hernquist,
  Astrophys.\ J.\  {\bf 471}, 115 (1996).

\bibitem{Silk:1997xw}
  J.~Silk and M.~J.~Rees,
  Astron.\ Astrophys.\  {\bf 331}, L1 (1998)
  [arXiv:astro-ph/9801013].

\bibitem{DiMatteo:2005sd}
  T.~Di Matteo, V.~Springel and L.~Hernquist,
  Nature {\bf 433}, 604 (2005)
  [arXiv:astro-ph/0502199].

\bibitem{Frank:1976uy}
  J.~Frank and M.~J.~a.~Rees,
  Mon.\ Not.\ Roy.\ Astron.\ Soc.\  {\bf 176}, 633 (1976).

\bibitem{Sigurdsson:2003ei}
  S.~Sigurdsson,
  Class.\ Quant.\ Grav.\  {\bf 20}, S45 (2003)
  [arXiv:astro-ph/0304251].

\bibitem{Sigurdsson:1996uz}
  S.~Sigurdsson and M.~J.~Rees,
  Mon.\ Not.\ Roy.\ Astron.\ Soc.\  {\bf 284}, 318 (1997)
  [arXiv:astro-ph/9608093].

\end{thebibliography}
\end{document}